\documentclass[useAMS,usenatbib]{mn2e}
\usepackage{mathptmx}
\usepackage{amssymb}
\usepackage{journals}
\usepackage{color} 

\usepackage{graphicx}
\usepackage[update,prepend]{epstopdf}
\usepackage{hyperref}
\usepackage[modulo]{lineno}

\newcommand{\Msun}{\ensuremath{{\rm M}_{\odot}}}

\title[Neutron-capture and Sc abundances of field halo stars]{Sc and
neutron-capture abundances in Galactic low- and high-$\alpha$ field halo
stars\thanks{This paper includes data gathered with the 6.5 meter Magellan
Telescopes located at Las Campanas Observatory, Chile.}}

\author[Fishlock et al.]{C. K.
Fishlock$^{1}$, D.
Yong$^{1}$\thanks{david.yong@anu.edu.au}, A. I.
Karakas$^{1,2}$,  A. Alves-Brito$^{3}$, J. Mel\'{e}ndez$^{4}$, \newauthor P. E.
Nissen$^{5}$, C. Kobayashi$^{6}$, and A. R. Casey$^{7}$ \\
$^{1}$Research School of Astronomy \& Astrophysics, Australian National
University, Canberra ACT 2611, Australia\\
$^{2}$ Monash Centre for Astrophysics, School of Physics and Astronomy, Monash
University, VIC 3800, Australia\\
$^{3}$ Instituto de Fisica, Universidade Federal do Rio Grande do Sul, Porto
Alegre, RS, Brazil\\
$^{4}$ Departamento de Astronomia do IAG/USP, Universidade de S\~ao Paulo,
Brazil \\
$^{5}$ Stellar Astrophysics Centre, Department of Physics and Astronomy, Aarhus
University, Ny Munkegade 120, DK-8000 Aarhus C, Denmark \\
$^{6}$ Centre for Astrophysics Research, Science and Technology Research
Institute, University of Hertfordshire, AL10 9AB, UK \\
$^{7}$ Institute of Astronomy, University of Cambridge, Cambridge, CB3 0HA, UK
}

\begin{document}

\date{Accepted .... Received ...; in original form ...}
\pagerange{\pageref{firstpage}--\pageref{lastpage}} \pubyear{2013}
\maketitle
\label{firstpage}

\begin{abstract}

We determine relative abundance ratios for the neutron-capture elements Zr, La,
Ce, Nd, and Eu for a sample of 27 Galactic dwarf stars with $-1.5 <$ [Fe/H]
$<-0.8$. We also measure the iron-peak element Sc. These stars separate into
three populations (low- and high-$\alpha$ halo and thick-disc stars) based on
the [$\alpha$/Fe] abundance ratio and their kinematics as discovered by Nissen
\& Schuster. We find differences between the low- and high-$\alpha$ groups in
the abundance ratios of [Sc/Fe], [Zr/Fe], [La/Zr], [Y/Eu], and [Ba/Eu] when
including Y and Ba from Nissen \& Schuster. For all ratios except [La/Zr], the
low-$\alpha$ stars have a lower abundance compared to the high-$\alpha$ stars.
The low-$\alpha$ stars display the same abundance patterns of high [Ba/Y] and
low [Y/Eu] as observed in present-day dwarf spheroidal galaxies, although with
smaller abundance differences, when compared to the high-$\alpha$ stars. These
distinct chemical patterns have been attributed to differences in the star
formation rate between the two populations and the contribution of
low-metallicity, low-mass asymptotic giant branch (AGB) stars to the
low-$\alpha$ population. By comparing the low-$\alpha$ population with AGB
stellar models, we place constraints on the mass range of the AGB stars. 

\end{abstract}


\section{Introduction}

Understanding the earliest phases of the assembly of our Galaxy, the Milky Way,
remains a key issue in contemporary astronomy.  Cosmological simulations
predict that mergers and accretion of satellite dwarf galaxies may play an
important role through hierarchical structure formation
\citep[e.g.,][]{Bullock:2005aa,Pillepich:2014aa}. Our Galaxy is no exception.
Observational support for the hierarchical nature of Galactic formation, and
the role dwarf galaxies play in that process, comes from the accretion of the
Sagittarius dwarf galaxy \citep{Ibata:1994aa} and numerous tidal streams
observed in the Galactic halo \citep[e.g.,][]{Belokurov:2006aa}. 

The chemical abundances and kinematics of Galactic halo stars hold valuable
clues concerning Galactic formation at the earliest times
\citep[e.g.,][]{Eggen:1962aa,Searle:1978aa,Venn:2004aa,Frebel:2015aa}. Due to
their long lifetimes, FGK-type dwarf stars are ideal for tracing the evolution
of the Galaxy. The atmospheres of dwarf stars closely mirror the chemical
composition of the gas clouds from which they formed. This means that
spectroscopic studies of these `stellar fossils' will enable us to probe the
chemical and kinematic history of our Galaxy over cosmic time. 

Comparison of chemical abundances between the Galactic halo and satellite dwarf
spheroidal galaxies (dSph) reveals differences in their chemical composition,
most notably in the [$\alpha$/Fe] ratio
\citep[e.g.,][]{Shetrone:2001aa,Tolstoy:2003aa,Venn:2004aa,Tolstoy:2009aa}. On
average, dSph stars with [Fe/H] $\gtrsim -2$ have lower [$\alpha$/Fe] than the
Galaxy. The [$\alpha$/Fe] ratio depends on the relative number of Type Ia to
Type II supernovae, and this quantity is sensitive to the star formation rate
\citep[e.g.,][]{Tinsley:1979aa,Matteucci:1986aa,Wheeler:1989aa,Nomoto:2013aa}.
The lower [$\alpha$/Fe] observed in dSph would therefore suggest a slower star
formation rate relative to the Galaxy. Differences in the [$\alpha$/Fe] ratio
between the halo and present-day dwarf galaxies preclude the continuous
merging of such systems although it may be that the mergers of massive dwarf
galaxies occurred at very early times in the formation of the Galactic halo
\citep{Venn:2004aa}.

Although a handful of $\alpha$-poor stars were known to exist in the Galactic
halo \citep[e.g.,][]{Carney:1997aa,King:1997aa,Preston:2000aa,Ivans:2003aa},
the study by \citet{Nissen:2010aa} (hereafter NS10) provided the first
compelling evidence that the halo hosts two populations that are well separated
in [$\alpha$/Fe]. NS10 studied 94 halo dwarf stars in the solar neighbourhood
using a careful differential analysis resulting in abundance errors for
[$\alpha$/Fe] of only 0.02~dex. They argued that the low-$\alpha$ stars have
kinematics consistent with accretion from a satellite galaxy with a slower star
formation rate than the high-$\alpha$ population which most likely formed
in-situ in the Galaxy. Additionally, NS10 found that the low- and high-$\alpha$
populations separate in the [Na/Fe] vs.\ [Ni/Fe] abundance plane. No separation
was seen for [Cr/Fe]. 

The study of NS10 was extended by \citet{Ramirez:2012aa}. Through an
analysis of the 777~nm O~{\sc I} triplet lines, they found that the
low-$\alpha$ stars have a lower [O/Fe] abundance compared to the high-$\alpha$
and thick-disc stars. The low-$\alpha$ stars also show a cosmic star-to-star
scatter in [O/Fe] at a given [Fe/H]. 

Before continuing, it is important to recognise that any attempts to infer the
chemical enrichment history of a stellar population requires a detailed
understanding of the nucleosynthetic origin of individual elements
\citep{Nissen:2013aa}. Low- to intermediate-mass stars undergo rich
nucleosynthesis once they reach the asymptotic giant branch (AGB) phase. It has
been observationally confirmed that the \emph{slow} neutron-capture process
($s$-process) takes place in AGB stars
\citep[e.g.,][]{Wallerstein:1997aa,Busso:1999aa}. The $s$-process is
responsible for the production of around half of the abundance of the heavy
elements beyond iron.  Through strong mass loss, AGB stars eject enriched
material into the interstellar medium and pollute the next generation of stars
\citep{Herwig:2005aa,Karakas:2014ab}.  

AGB stars may play an important role in the origin of chemical abundance
differences between the Galaxy and nearby dwarf galaxies. In addition to
[$\alpha$/Fe], differences in the ratios of [Ba/Y] and [Y/Eu] are found between
dSph and the Galaxy (e.g., \citealt{Tolstoy:2009aa}). The high [Ba/Y] observed
in dSph was attributed to low-mass AGB stars enriching the interstellar medium
as a result of the slower chemical evolution. To explain the low [Y/Eu]
observed, \citet{Venn:2004aa} suggested that metal-poor AGB stars did not
contribute as much Y (compared to more metal-rich AGB stars) because the
production of second $s$-process peak elements such as Ba and La were favoured
over first $s$-process peak elements such as Zr and Y \citep{Busso:2001aa}.  

\citet{Nissen:2011aa} (hereafter NS11) determined abundances for two
neutron-capture elements produced by the $s$-process in AGB stars, Ba and Y
(along with the iron-peak elements Mn, Cu, and Zn). They discovered that the
low- and high-$\alpha$ populations could be separated in [Cu/Fe] and [Zn/Fe].
More importantly, NS11 found that the [Ba/Y] ratio differs between the two
populations with the low-$\alpha$ stars showing a higher [Ba/Y] value. Those
chemical abundance patterns are similar to that found in dSph by
\citet{Venn:2004aa} and reinforces the hypothesis that the low-$\alpha$ stars
are likely to have been accreted from dSph. 

The NS10 sample provides a unique opportunity to study additional
neutron-capture element abundances in the low- and high-$\alpha$ populations.
Such an analysis will enable us to examine the abundances for a larger range of
neutron-capture elements, with high-precision, in stars likely to have been
accreted from dwarf galaxies. Stars currently in dwarf galaxies are very faint
such that high precision chemical abundance studies cannot be conducted with
existing facilities. The aim of this study is to investigate the role of AGB
stars in the chemical enrichment of the Galactic halo and dSph.  Specifically,
we build upon the work of NS10 and NS11 by studying additional $s$-process
elements (Zr, La, Ce, and Nd). We also measure, for the first time with this
unique sample, a \emph{rapid} neutron-capture process ($r$-process) element
(Eu). Finally, we measure Sc which has been observed to behave like an $\alpha$
element \citep[e.g.,][]{Nissen:2000ab}. 

\section{Sample selection and observations}

Our observations focused on obtaining high quality blue spectra for stars in
the NS10 sample. In the analyses of NS10 and NS11, only lines redder than
4700~\AA~were measured. Our targets were selected to include both low- and
high-$\alpha$ stars across the metallicity range of $-1.5 <$ [Fe/H] $< -0.8$.
The sample consists of 13 low-$\alpha$ stars, 6 high-$\alpha$ stars, and 8
thick-disc (TD) stars, including the two reference TD stars used in the studies
by NS10 and NS11, HD 22879 and HD 76932. 

Programme stars were observed using the Magellan Inamori Kyocera Echelle (MIKE)
spectrograph \citep{Bernstein:2003aa} on the Magellan (Clay) telescope 
during February 2011 and June 2011. We used the 0.35'' slit size which provides
a resolving power of 83,000 in the blue. Typical exposure times ranged from 500
to 5000 seconds per target and the signal-to-noise ratio per pixel of the
spectra was approximately 400  at 4500~\AA. The wavelength coverage provided by
MIKE is 3350 to 9000~\AA.  Data reduction was accomplished using {\sc
iraf}\footnote{{\sc iraf} is distributed by the National Optical Astronomy
Observatory, which is operated by the Association of Universities for Research
in Astronomy (AURA) under a cooperative agreement with the National Science
Foundation.} with the MIKE {\sc
mtools}\footnote{\url{www.lco.cl/telescopes-information/magellan/instruments/mike/iraf-tools/iraf-mtools-package}}
package. Continuum fitting was performed using the Spectroscopy Made Hard ({\sc
smh}) package as used in \citet{Casey:2014ab} and described in
\citet{Casey:2014aa}. 

\section{Abundances}

Following NS10, we perform a differential line-by-line analysis relative to the
TD star HD 22879. This approach enables us to obtain high-precision abundance
ratios which are necessary in identifying possible separations between the low-
and high-$\alpha$ populations. The model atmospheres used in the analysis were
one-dimensional, plane parallel, local thermodynamic equilibrium (LTE) {\sc
atlas9} models by \citet{Castelli:2003aa}. Interpolation between $T_{\rm eff}$,
$\log g$, [Fe/H], and [$\alpha$/Fe] was required to produce model atmospheres.
The stellar parameters, $T_{\rm eff}$,  $\xi_{\rm turb}$, and log $g$, as well
as the abundance ratios of [Fe/H] and [$\alpha$/Fe] were taken from NS10 and
NS11 (see Table~\ref{tab:params}). For a complete discussion on the
determination of these parameters see NS10 and NS11. We use stellar parameters
from NS10 without the correction for effective temperature of $\sim$ 100~K from
NS11. Given the differential nature of the abundance analysis and that most
stars have fairly similar stellar parameters, a systematic shift of 100~K is
likely to have a minimal effect on the final differential abundances.

\begin{table}
 \begin{center}
  \caption{Atmospheric parameters and abundance ratios for the programme stars (taken from NS10 and NS11).
 \label{tab:params}}
  \vspace{1mm}
   \begin{tabular}{lcccrrc}
\hline \hline
ID & $T_{\rm eff}$ & log $g$ & $\xi_{\rm turb}$ & [Fe/H] & [$\alpha$/Fe] & Class$^{a}$ \\
 & (K) & (cgs) & (km s$^{-1}$) &  & & \\ 
\hline
CD -45 3283 & 5597 & 4.55 & 1.0 & $-$0.91 & 0.12 & low-$\alpha$  \\
CD -61 0282 & 5759 & 4.31 & 1.3 & $-$1.23 & 0.22 & low-$\alpha$  \\ 
G 16-20 & 5625 & 3.64 & 1.5 & $-$1.42 & 0.26 & (low-$\alpha$)  \\
G 18-28 & 5372 & 4.41 & 1.0 & $-$0.83 & 0.31 & high-$\alpha$  \\
G 18-39 & 6040 & 4.21 & 1.5 & $-$1.39 & 0.34 & high-$\alpha$  \\
G 31-55 & 5638 & 4.30 & 1.4 & $-$1.10 & 0.29 & high-$\alpha$  \\
G 56-30 & 5830 & 4.26 & 1.3 & $-$0.89 & 0.11 & low-$\alpha$  \\
G 56-36 & 5933 & 4.28 & 1.4 & $-$0.94 & 0.20 & low-$\alpha$  \\
G 66-22 & 5236 & 4.41 & 0.9 & $-$0.86 & 0.12 & low-$\alpha$  \\
G 170-56 & 5994 & 4.12 & 1.5 & $-$0.92 & 0.17 & low-$\alpha$  \\
HD 3567 & 6051 & 4.02 & 1.5 & $-$1.16 & 0.21 & low-$\alpha$  \\
HD 22879 & 5759 & 4.25 & 1.3 & $-$0.85 & 0.31 & TD  \\
HD 25704 & 5868 & 4.26 & 1.4 & $-$0.85 & 0.24 & TD  \\
HD 76932 & 5877 & 4.13 & 1.4 & $-$0.87 & 0.29 & TD  \\
HD 103723 & 5938 & 4.19 & 1.2 & $-$0.80 & 0.14 & low-$\alpha$  \\
HD 105004 & 5754 & 4.30 & 1.2 & $-$0.82 & 0.14 & low-$\alpha$  \\
HD 111980 & 5778 & 3.96 & 1.5 & $-$1.08 & 0.34 & high-$\alpha$  \\
HD 120559 & 5412 & 4.50 & 1.1 & $-$0.89 & 0.30 & TD  \\
HD 126681 & 5507 & 4.45 & 1.2 & $-$1.17 & 0.35 & TD  \\
HD 179626 & 5850 & 4.13 & 1.6 & $-$1.00 & 0.32 & high-$\alpha$  \\
HD 189558 & 5617 & 3.80 & 1.4 & $-$1.12 & 0.35 & TD  \\
HD 193901 & 5650 & 4.36 & 1.2 & $-$1.07 & 0.17 & low-$\alpha$  \\
HD 194598 & 5942 & 4.33 & 1.4 & $-$1.09 & 0.20 & low-$\alpha$  \\
HD 199289 & 5810 & 4.28 & 1.3 & $-$1.04 & 0.30 & TD  \\
HD 205650 & 5698 & 4.32 & 1.3 & $-$1.17 & 0.30 & TD  \\
HD 219617 & 5862 & 4.28 & 1.5 & $-$1.45 & 0.28 & (low-$\alpha$)  \\
HD 230409 & 5318 & 4.54 & 1.1 & $-$0.85 & 0.27 & high-$\alpha$   \\
\hline \hline
  \end{tabular} 
\\ $^a$For stars with [Fe/H] $< -$1.4, the classification is uncertain (NS10).
 \end{center}
\end{table}

\begin{figure}
\begin{center}
\includegraphics[]{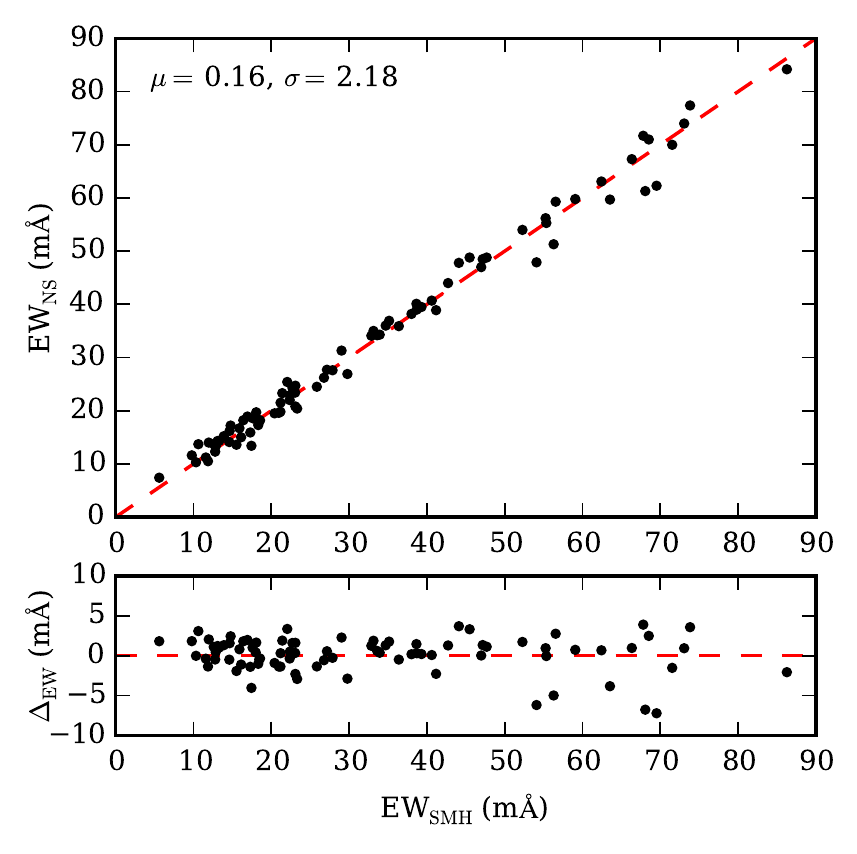} 
\caption{Comparison between EW measurements for HD 22879 from NS10 and NS11
(EW$_{\rm NS}$) and by using {\sc smh} (EW$_{\rm SMH}$, this work). The dashed
line shows the one-to-one correspondence line. The bottom panel shows the
difference between the two analyses.}
\label{fig:ew}
\end{center}
\end{figure}

Elemental abundances for Zr, Ce, and Nd were derived from the analysis of
equivalent widths (EWs) using {\sc smh}. For lines with EW $<$ 60~m\AA, a
Gaussian profile was fitted to the line. For lines with EW larger than 60~m\AA,
a Voigt profile was used.  Figure~\ref{fig:ew} presents a comparison of 85 EW
measurements (excluding Fe I and Fe II) of NS10 and NS11 to the EWs measured
using {\sc smh} with the same line list for HD 22879. The mean EW difference
between the two methods is 0.16~$\pm$~0.24~m\AA\ ($\sigma$ = 2.18~m\AA). 

\begin{table}
 \begin{center}
  \caption{EW measurements (in m\AA) of Zr II, Ce II, and Nd II for each star.
 \label{tab:equiv}}
  \vspace{1mm}
   \begin{tabular}{lcccc}
\hline \hline
ID & Zr II & Ce II & Ce II & Nd II \\
 & 4208.98~\AA & 4562.36~\AA & 4628.16~\AA & 4706.54~\AA \\ 
\hline
CD -45 3283 & 21.3 & 7.6 & 5.3 & 7.0   \\
CD -61 0282 & 17.7 & 5.2 & 2.7 & 4.0   \\
G 16-20 & 22.1 & 6.6 & 4.1 & 4.0   \\
G 18-28 & 29.1 & 10.3 & \dots & 9.4   \\
G 18-39 & 17.3 & \dots & 2.0 & 3.0   \\
G 31-55 & 23.8 & 4.9 & 3.6 & 4.6   \\
G 56-30 & 23.1 & 6.6 & 7.2 & 2.9   \\
G 56-36 & 26.1 & 6.7 & 4.4 & 5.4   \\
G 66-22 & 26.2 & \dots & 8.9 & 10.6   \\
G 170-56 & 23.4 & 6.8 & 3.4 & 4.2   \\
HD 3567 & 21.3 & 4.9 & 5.3 & 4.0   \\
HD 22879 & 31.4 & 8.2 & 6.5 & 7.9   \\
HD 25704 & 22.4 & 6.5 & 6.1 & 7.1   \\
HD 76932 & 34.3 & 8.4 & 5.1 & 7.9   \\
HD 103723 & 27.5 & 9.4 & 7.4 & 7.7   \\
HD 105004 & 24.6 & \dots & 9.4 & 5.0   \\
HD 111980 & 37.4 & 10.6 & 6.5 & 6.6   \\
HD 120559 & 23.8 & \dots & 4.1 & 7.5   \\
HD 126681 & 28.9 & 7.4 & 5.4 & 5.0   \\
HD 179626 & 27.6 & 5.9 & 4.1 & 4.1   \\
HD 189558 & 39.3 & 10.9 & 8.9 & 7.1   \\
HD 193901 & 17.4 & 5.2 & \dots & 5.3   \\
HD 194598 & 15.8 & 3.6 & 3.2 & 2.9   \\
HD 199289 & 19.1 & 3.3 & 2.5 & 3.6   \\
HD 205650 & 19.1 & 4.6 & 3.4 & 3.0   \\
HD 219617 & 10.7 & \dots & 3.3 & 2.0$^{a}$   \\
HD 230409 & 26.2 & 8.2 & 5.8 & 10.5   \\
\hline \hline
  \end{tabular} 
\\ $^a$ Upper limit
 \end{center}
\end{table}

We use the atomic data compiled by \citet{Yong:2005aa} for the measured
absorption lines.  The abundance of Zr is derived from the EW of the Zr II line
at 4208.98~\AA~which is unblended with other lines, allowing for a reliable
measurement. The Zr II line is detected in all stars with all EW measurements
being greater than 10~m\AA~(see Table~\ref{tab:equiv}).  Ce abundances have
been determined using the EWs of two Ce II lines at 4562.36~\AA~and
4628.16~\AA. Seven stars only have one Ce II line available for measurement: 
five stars do not have a measurable 4562.36~\AA~line whereas two stars do not
have a measurable 4628.16~\AA~line (see Table~\ref{tab:equiv}). For all but
three Ce lines, the EW is less than 10 m\AA{}.

Nd abundances have been determined using the Nd II line at 4706.54 \AA{}. The
EW measurements are less than 10 m\AA{} except in two stars (G 66-22 and HD
230409). We arbitrarily set a minimum reliable EW measurement limit of 2 m\AA.
One star, the most metal-poor in the sample (HD 219617), falls below our 2
m\AA{} limit and therefore we adopt an upper limit of 2 m\AA{} for the Nd EW.

\begin{figure}
\begin{center}
\includegraphics[]{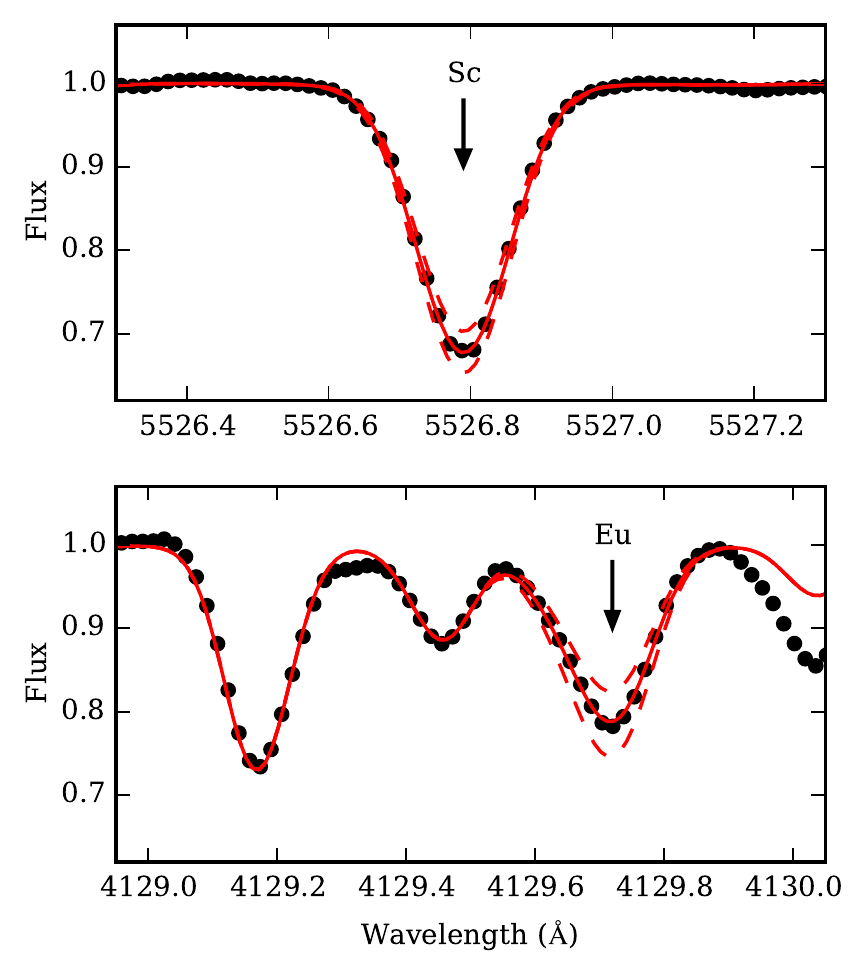} 
\caption{Synthetic spectrum fit (solid line) to the observations (black points)
of the Sc line at 5526~\AA~(top) and the  Eu line at 4129~\AA~(bottom) for HD
22879. The dashed lines show synthetic spectra with [Sc/Fe] and [Eu/Fe] of $\pm
0.1.$}
\label{fig:synth}
\end{center}
\end{figure}

For Sc, La, and Eu, we fit synthetic spectra generated by the LTE spectrum
synthesis program MOOG \citep[v. 2013,][]{Sneden:1973aa} to determine
abundances. These elements require the consideration of hyperfine splitting
(HFS) and/or isotopic shifts. Figure~\ref{fig:synth} illustrates the fit of the
synthetic spectrum to the observed spectrum for the Sc and Eu lines of HD
22879. The best fitting synthetic spectrum was determined using $\chi^2$
minimisation. For Sc, the log $gf$ values of \citet{Nissen:2000ab} were
utilised for the 5526.81~\AA~absorption line. La was determined using one La II
line at 4086.71~\AA~with log $gf$ values from \citet{Lawler:2001aa}. For the Eu
II spectral line at 4129.72~\AA~we use the log $gf$ values from
\citet{Lawler:2001ab} and solar isotope ratios from \citet{Lodders:2003aa}.

We adopt solar elemental abundances from \citet{Asplund:2009aa} as detailed in
Table~\ref{tab:solar_abund}. Since we perform a differential analysis to measure
abundances, the choice of solar abundance only shifts the stellar abundances up
or down and does not affect the relative abundances. To reiterate, we are
primarily interested in relative abundances, and therefore abundance
differences, between the low- and high-$\alpha$ populations. The abundances are
given in Table~\ref{tab:abund}. 

\begin{table}
 \begin{center}
  \caption{Adopted solar elemental abundances from \citet{Asplund:2009aa}.
 \label{tab:solar_abund}}
  \vspace{1mm}
   \begin{tabular}{ccc}
\hline \hline
Element & Z & $\log(\rm{ X/H}) + 12$ \\
\hline
Sc$^{a}$ & 21 & 3.05 \\
Zr & 40 & 2.58  \\
La & 57 & 1.10  \\
Ce & 58 & 1.58 \\
Nd & 60 & 1.42  \\
Eu & 63 & 0.52 \\
\hline \hline
  \end{tabular} 
\\ $^a$ Meteoritic value.
 \end{center}
\end{table}

We estimated errors using the stellar parameter uncertainties from NS10 where
the differential errors for $\sigma(\log g)$ and $\sigma(T_{\rm eff})$ are $\pm
0.05$ and $\pm 30$~K, respectively. The 1-$\sigma$ errors from NS10 for [Fe/H]
and [$\alpha$/Fe] were 0.04 and 0.02 dex, respectively. We estimated the
1-$\sigma$ errors in [Zr/Fe] and [Ce/Fe] to be 0.04. For [Sc/Fe], [La/Fe],
and [Eu/Fe], the errors are 0.05. The error for [Nd/Fe] is 0.07.

\begin{table*}
 \begin{center}
  \caption{Abundances for each of the measured elements. The values for [Fe/H] are taken from NS10.
 \label{tab:abund}}
  \vspace{1mm}
   \begin{tabular}{lrrrrrrr}
\hline \hline
Star & [Fe/H] & [Sc/Fe] & [Zr/Fe] & [La/Fe] & [Ce/Fe] & [Nd/Fe] & [Eu/Fe] \\
\hline
CD -45 3283 & $-$0.91 & $-$0.11 & $-$0.07 & 0.00 & $-$0.01 & 0.36 & 0.31  \\
CD -61 0282 & $-$1.23 & $-$0.15 & 0.00 & $-$0.01 & $-$0.05 & 0.32 & 0.26  \\
G 16-20 & $-$1.42 & $-$0.11 & $-$0.00 & 0.02 & $-$0.04 & 0.17 & 0.32  \\
G 18-28 & $-$0.83 & 0.11 & 0.03 & $-$0.02 & 0.04 & 0.34 & 0.13  \\
G 18-39 & $-$1.39 & $-$0.00 & 0.21 & 0.07 & $-$0.03 & 0.47 & 0.26  \\
G 31-55 & $-$1.10 & $-$0.03 & 0.06 & $-$0.08 & $-$0.11 & 0.26 & 0.05  \\
G 56-30 & $-$0.89 & $-$0.22 & $-$0.12 & $-$0.06 & $-$0.04 & $-$0.09 & 0.09  \\
G 56-36 & $-$0.94 & $-$0.02 & 0.04 & $-$0.04 & $-$0.05 & 0.31 & 0.13  \\
G 66-22 & $-$0.86 & $-$0.16 & $-$0.08 & $-$0.04 & 0.02 & 0.34 & 0.30  \\
G 170-56 & $-$0.92 & $-$0.14 & $-$0.10 & $-$0.02 & $-$0.16 & 0.14 & 0.16  \\
HD 3567 & $-$1.16 & $-$0.04 & 0.05 & 0.16 & 0.07 & 0.33 & 0.39  \\
HD 22879 & $-$0.85 & 0.12 & 0.06 & $-$0.09 & $-$0.04 & 0.34 & 0.16  \\
HD 25704 & $-$0.85 & 0.09 & $-$0.16 & $-$0.02 & $-$0.09 & 0.33 & 0.15  \\
HD 76932 & $-$0.87 & 0.15 & 0.13 & 0.03 & $-$0.08 & 0.36 & 0.21  \\
HD 103723 & $-$0.80 & $-$0.03 & $-$0.05 & $-$0.03 & 0.01 & 0.33 & 0.17  \\
HD 105004 & $-$0.82 & $-$0.04 & $-$0.13 & $-$0.12 & 0.10 & 0.10 & 0.04  \\
HD 111980 & $-$1.08 & 0.00 & 0.25 & 0.13 & 0.09 & 0.33 & 0.14  \\
HD 120559 & $-$0.89 & 0.17 & $-$0.05 & $-$0.05 & $-$0.22 & 0.32 & 0.20  \\
HD 126681 & $-$1.17 & 0.00 & 0.28 & 0.14 & 0.13 & 0.34 & 0.20  \\
HD 179626 & $-$1.00 & 0.12 & 0.03 & 0.02 & $-$0.14 & 0.14 & 0.03  \\
HD 189558 & $-$1.12 & 0.09 & 0.24 & 0.07 & 0.08 & 0.24 & 0.17  \\
HD 193901 & $-$1.07 & $-$0.19 & $-$0.14 & $-$0.00 & $-$0.08 & 0.31 & 0.24  \\
HD 194598 & $-$1.09 & $-$0.07 & $-$0.11 & $-$0.03 & $-$0.11 & 0.17 & 0.27  \\
HD 199289 & $-$1.04 & 0.08 & $-$0.09 & $-$0.12 & $-$0.30 & 0.15 & 0.09  \\
HD 205650 & $-$1.17 & $-$0.01 & 0.01 & $-$0.03 & $-$0.07 & 0.14 & 0.19  \\
HD 219617 & $-$1.45 & $-$0.11 & $-$0.05 & $-$0.11 & 0.18 & 0.03$^{a}$ & 0.21  \\
HD 230409 & $-$0.85 & 0.05 & $-$0.03 & $-$0.00 & $-$0.07 & 0.43 & 0.27  \\
\hline \hline
  \end{tabular} 
\\ $^a$ Upper limit
 \end{center}
\end{table*}

\begin{figure}
\begin{center}
\includegraphics[]{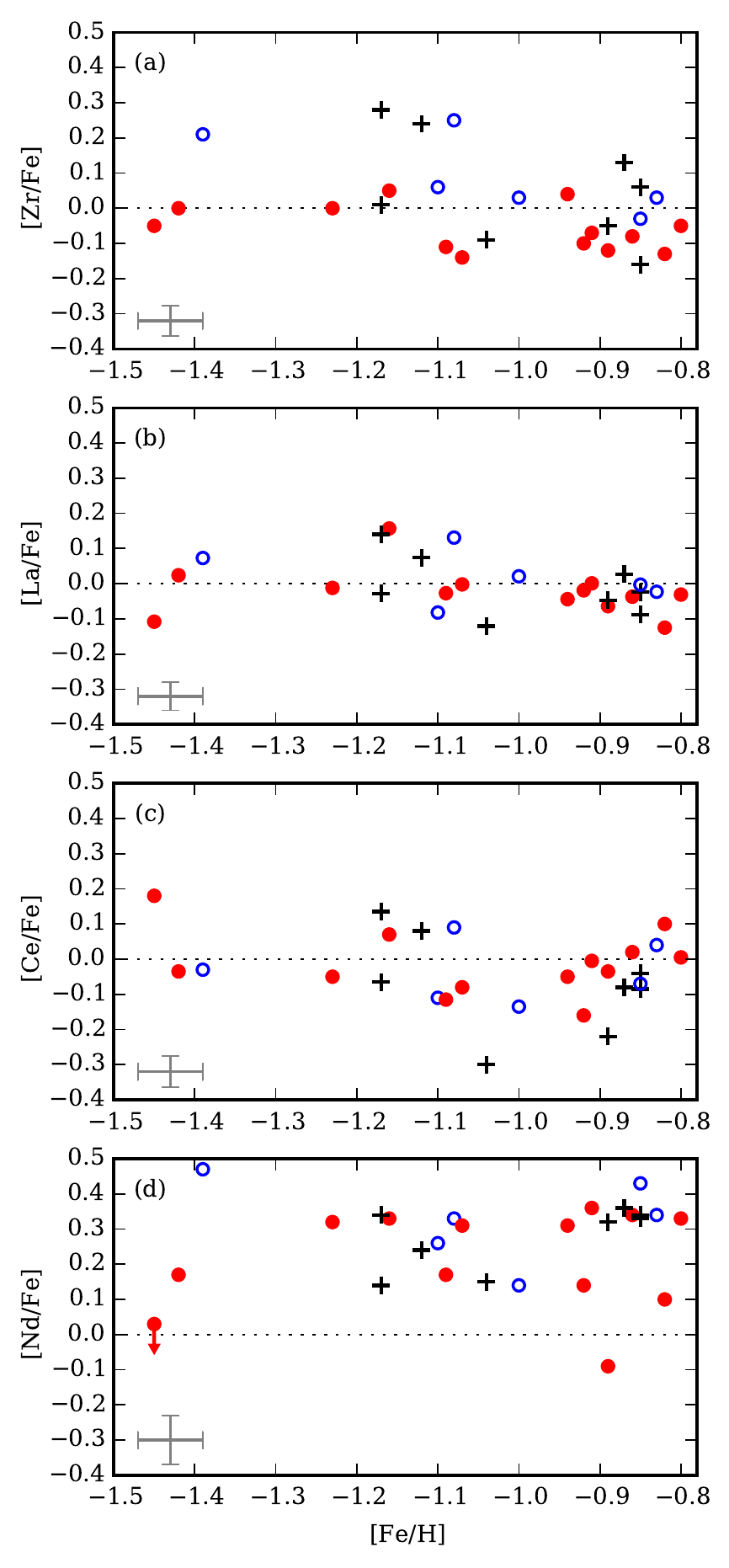} 
\caption{Four $s$-process elements (Zr, La, Ce, and Nd) versus [Fe/H]. The red
circles indicate low-$\alpha$ stars, the blue open circles indicate
high-$\alpha$ stars, and the plus signs indicate TD stars. The 1-$\sigma$ error
is shown in the bottom-left of each plot.}
\label{fig:zrlacend_fe}
\end{center}
\end{figure}

\subsection{Zr}

The neutron-capture element Zr is predominately produced by AGB stars and is a
first $s$-process peak element. The $s$-process contribution to Zr in the Solar
System is $\sim$80\%
\citep[e.g.,][]{Simmerer:2004aa,Sneden:2008aa,Bisterzo:2014aa}. By determining
the abundance of Zr, we expand on the measurements of Y, another first
s-process peak element whose abundances were previously obtained by NS11. 

Figure~\ref{fig:zrlacend_fe}(a) shows [Zr/Fe] as a function of [Fe/H]. The
dispersion in [Zr/Fe] is $\sim 0.12$ dex which is larger than the error of 0.04
dex. When considering the two populations, the scatter in [Zr/Fe] is smaller
for the low-$\alpha$ stars with a dispersion of 0.06 dex compared to a
dispersion of 0.13 dex when combining the high-$\alpha$ and TD stars. In
addition to the difference in the dispersion of [Zr/Fe], the low-$\alpha$ stars
have a systematically lower [Zr/Fe] abundance compared to the high-$\alpha$ and
TD stars. 

\begin{figure}
\begin{center}
\includegraphics[]{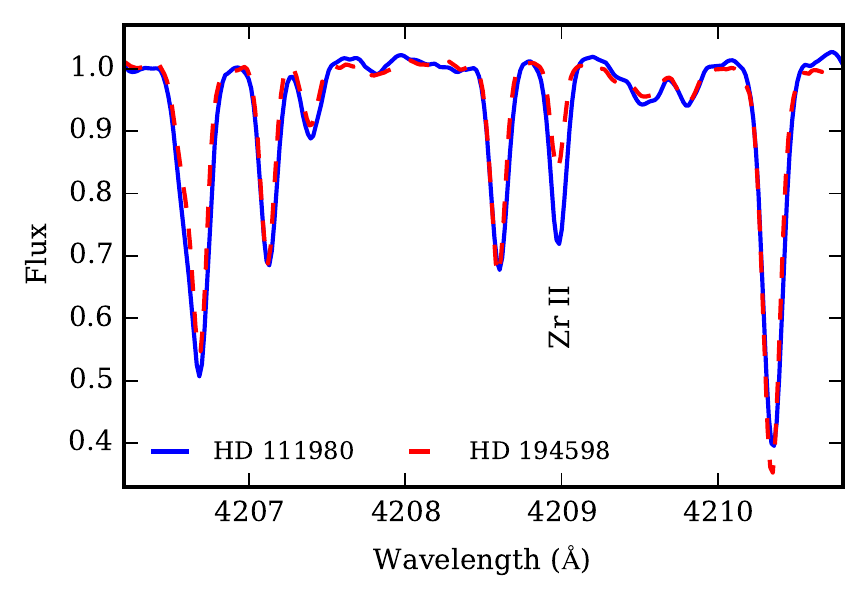} 
\caption{MIKE spectra around the Zr line at 4208.98~\AA~for a low-$\alpha$ star
(HD 194598) and a high-$\alpha$ star (HD 111980).}
\label{fig:spectra_zr}
\end{center}
\end{figure}

This possible separation in the high- and low-$\alpha$ populations is in
agreement with the slight separation in abundance measurements of [Y/Fe] for
the two populations seen in NS11. As Zr and Y are both first-peak $s$-process
elements, they are expected to follow the same trend.
Figure~\ref{fig:spectra_zr} demonstrates the difference in the Zr abundance for
a low-$\alpha$ star (HD 194598) and a high-$\alpha$ star (HD 111980) at nearly
the same metallicity (see Table~\ref{tab:params}). The low-$\alpha$ star has a
weaker Zr II line than the high-$\alpha$ star. While the effective temperature
differs between these two stars by approximately 150~K, all other lines in this
region exhibit very similar strengths. 

\begin{figure}
\begin{center}
\includegraphics[]{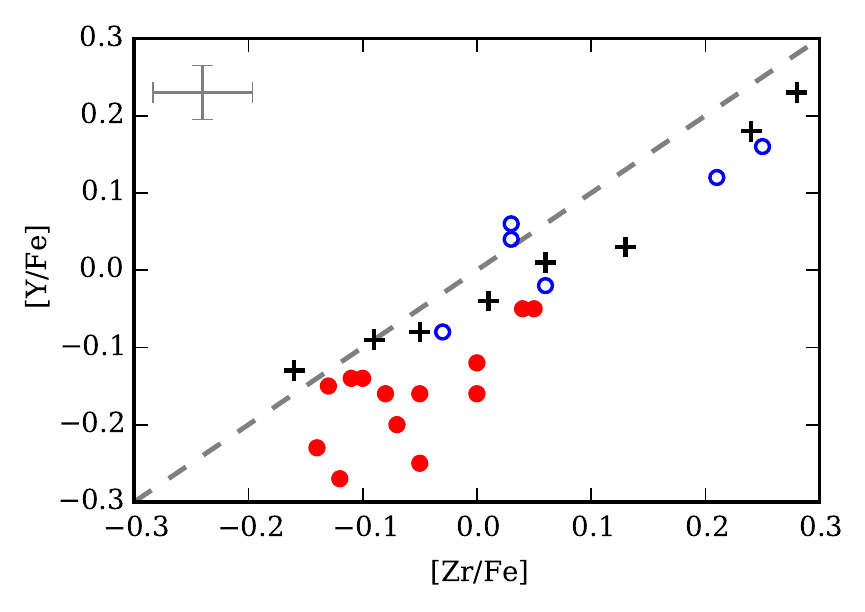} 
\caption{Comparison of [Y/Fe] (taken from NS11) and [Zr/Fe]. The red circles
indicate low-$\alpha$ stars, the blue open circles indicate high-$\alpha$
stars, and the plus signs indicate TD stars. The dashed line shows the
one-to-one correspondence line. The 1-$\sigma$ error is shown in the
top-left.}
\label{fig:zr_y}
\end{center}
\end{figure}

A comparison between [Y/Fe] as measured by NS11 and [Zr/Fe] is presented in
Figure~\ref{fig:zr_y}. The abundance of Zr follows the abundance of Y which is
expected if Y and Zr have a common origin. While there is a systematic offset
between [Y/Fe] and [Zr/Fe] which may be due to errors in the $gf$ values of the
lines, the important point is that the slope of the data is close to 1. The
high-$\alpha$ stars have a tighter correlation between [Y/Fe] and [Zr/Fe]
compared to the low-$\alpha$ stars. 

\subsection{La, Ce, and Nd}

We extend upon the Ba measurements of NS11 by determining abundances for other
elements belonging to the second $s$-process peak, namely La, Ce, and Nd.

The $s$-process contribution to La in the Solar System is 75\% and there are a
number of advantages with using La as an $s$-process indicator compared to Ba
(e.g., \citealt{Simmerer:2004aa,Winckler:2006aa}). While Ba consists of
five stable isotopes and requires the consideration of isotope shifts, 99.91\%
of La is in the isotope $^{139}$La \citep{Lodders:2003aa,Asplund:2009aa}.
Figure~\ref{fig:zrlacend_fe}(b) presents [La/Fe] as a function of [Fe/H]. There
is no clear difference in the La abundance between the low-$\alpha$,
high-$\alpha$, and TD stars. The dispersion in the abundance of [La/Fe] is
$\sim 0.07$ dex, the lowest of all the $s$-process abundances measured here.
The [La/Fe] ratio is almost constant with [Fe/H] and the mean abundance close
to the solar value. 

The $s$-process contribution to Ce in the Solar System is 81\% where
$^{140}$Ce is the only stable isotope accessible by the $s$-process
\citep{Sneden:2008aa}.  Figure~\ref{fig:zrlacend_fe}(c) presents the
results for [Ce/Fe]. The majority of stars have a subsolar [Ce/Fe] abundance
with a mean value of around $-$0.04 dex and no apparent separation between the
low- and high-$\alpha$ stars. The dispersions for the low-$\alpha$ and
high-$\alpha$ stars are essentially the same (0.09 dex compared to 0.10 dex,
respectively). 

For Nd, 58\% of the Solar System abundance is due to the $s$-process
\citep{Sneden:2008aa}. Figure~\ref{fig:zrlacend_fe}(d) presents the abundance
of [Nd/Fe] against [Fe/H]. Due to the weak Nd II line for HD 219617, we provide
an upper limit of [Nd/Fe] $= 0.03$ for this object based on an EW limit of
2~m\AA.  There is no obvious difference in [Nd/Fe] between the two populations
except for a few low-$\alpha$ stars having a lower Nd abundance. This causes
the dispersion to be higher for the low-$\alpha$ stars relative to the
high-$\alpha$ stars, 0.14 dex compared 0.10 dex. The mean abundance of [Nd/Fe]
is 0.26 dex and all but one star has a supersolar [Nd/Fe] abundance. 

\subsection{Eu}

Eu is predominately an $r$-process element with less than 2\% of the
Solar System abundance a result of $s$-process nucleosynthesis
\citep{Sneden:2008aa}. We have included Eu in this study to investigate
the  contribution of $r$-process nucleosynthesis to the low- and high-$\alpha$
stars. 

\begin{figure}
\begin{center}
\includegraphics[]{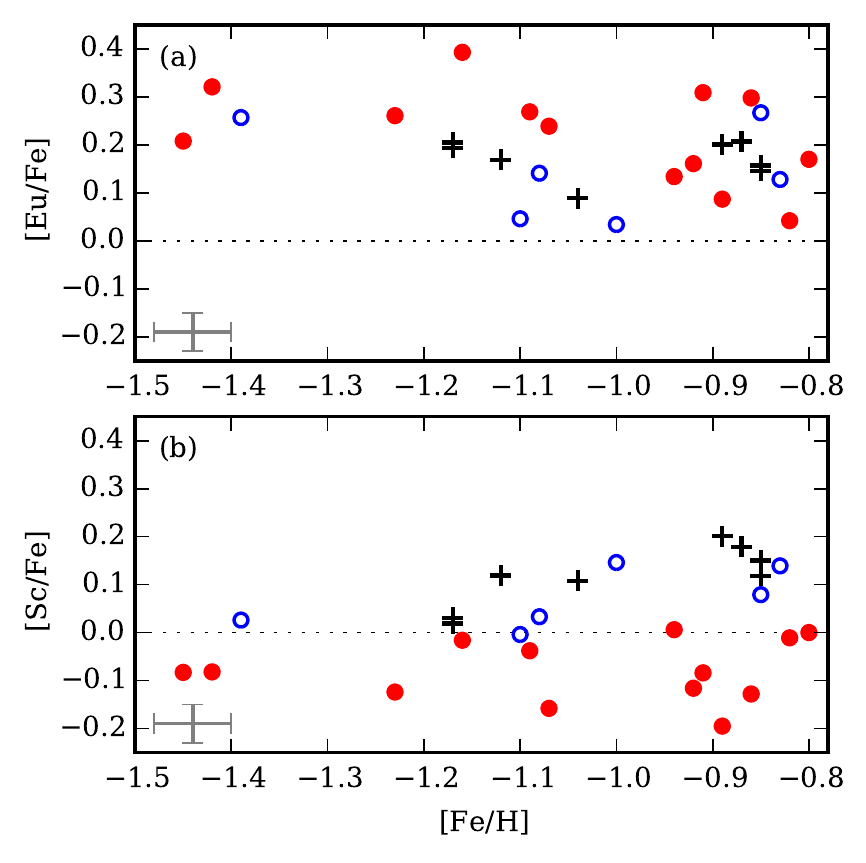} 
\caption{($a$) [Sc/Fe] and ($b$) [Eu/Fe] versus [Fe/H]. The red circles
indicate low-$\alpha$ stars, the blue open circles indicate high-$\alpha$
stars, and the plus signs indicate TD stars. The 1-$\sigma$ error is shown in
the bottom-left of each plot.}
\label{fig:sceu_fe}
\end{center}
\end{figure}

Figure~\ref{fig:sceu_fe}(a) presents the [Eu/Fe] abundances where there is a
potential separation between the two populations: the low-$\alpha$ stars appear
to show a higher [Eu/Fe] abundance compared to the high-$\alpha$ stars.
However, due to the lack of stars at lower metallicities, the separation is not
as clear as what is found for other elements such as Zr. The low-$\alpha$ group
also shows a potential decrease in [Eu/Fe] with increasing metallicity. A
larger sample of stars is needed to confirm these results. 

\subsection{Sc}

There is only one stable isotope of Sc ($^{45}$Sc) and it is mainly synthesised
during carbon and neon burning in massive stars \citep{Woosley:2002aa}. It is
also produced to a lesser extent in AGB stars \citep[e.g.,][]{Fishlock:2014aa}.
We measure Sc in order to verify the conclusions of \citet{Nissen:2000ab} in
which the low-$\alpha$ stars have a lower [Sc/Fe] abundance compared to the
high-$\alpha$ stars. 

Figure~\ref{fig:sceu_fe}(b) reveals a clear separation in [Sc/Fe]. The
low-$\alpha$ stars have a subsolar [Sc/Fe] abundance with a mean value of
$-$0.11 dex whereas the high-$\alpha$ stars have a mostly supersolar abundance
with a mean value of 0.07 dex. These results confirm the separation in [Sc/Fe]
found in the earlier work by \citet{Nissen:2000ab} between the low- and
high-$\alpha$ populations. 

\section{Analysis}

As with the NS10 and NS11 studies, we combine the TD and high-$\alpha$ halo
stars into one high-$\alpha$ group. Due to a smaller sample size compared to
NS11, we do not remove the abundance dependence on [Fe/H] for the high-$\alpha$
stars as it cannot be definitively determined. Given that our sample size is
smaller than in NS10, we now seek to quantify this potential limitation by
introducing our methodology on how we decide whether a given abundance ratio
exhibits a difference (i.e., separation) between the low- and high-$\alpha$
groups. 

We first consider the NS10 measurements of Cr, a representative element for
which NS10 found no significant separation between the low- and high-$\alpha$
groups. Using a two sample Kolmogorov-Smirnov (KS) test between the low- and
high-$\alpha$ groups (without distinguishing TD stars), we find a $p$-value of
0.07 when considering [Cr/Fe] for all stars in the NS10 sample with $-1.45
\leq$ [Fe/H] $\leq -0.8$. 

In order to investigate how our smaller sample size affects our ability to
identify abundance separations, we performed the following tests. We randomly
select 13 low-$\alpha$ and 14 high-$\alpha$ stars from the full NS10 sample
with $-1.45 \leq$ [Fe/H] $\leq -0.8$, apply the KS test and obtain the
$p$-value. We repeat this exercise 10,000 times and find that for approximately
25\% of the realisations, the $p$-value is below 0.07. If we arbitrarily define
the threshold to be a $p$-value of 0.03, roughly 10\% of the random
realisations result in a $p$-value below 0.03. 

We now examine [Ba/Y], an abundance ratio that was found to exhibit a
significant separation between the low- and high-$\alpha$ populations. For the
full NS10 sample with $-1.45 \leq$ [Fe/H] $\leq -0.8$, we compute a $p$-value
of $2.9 \times 10^{-9}$. We then seek to quantify the effect of having a
smaller sample size. As above, we randomly select 13 low-$\alpha$ and 14
high-$\alpha$ stars from the full NS10 sample with $-1.45 \leq$ [Fe/H] $\leq
-0.8$, apply the KS test and obtain the $p$-value. We repeat the exercise 10,000
times. For a threshold $p$-value of 0.03, about 99.7\% of random
realisations have a $p$-value below this level, indicating that we are able to
identify the separation in [Ba/Y] in all but 0.3\% of cases.

Based on these tests, we choose a $p$-value of 0.03 which ensures that there
is at least a 90\% chance that the abundance separation in our sample of
27 stars is genuine despite having a smaller number of stars than NS10 and
NS11. We also confirm using a KS test that the high-$\alpha$ population and TD
stars have a significant probability of coming from the same population. In
Table~\ref{tab:med_std} we present the mean of each abundance ratio and its
standard deviation for both low- and high-$\alpha$ groups. We also present the
$p$-value of the two sample KS test between the low- and high-$\alpha$ groups.
The analyses of [Y/Fe], [Ba/Fe], and [Ba/Y]  are included in
Table~\ref{tab:med_std} for comparison using the abundances measured by NS11
for our sample of 27 stars. For comparison, the mean $p$-value calculated
for [Mg/Fe] when randomly selecting 13 low-$\alpha$ and 14 high-$\alpha$ stars
is $4.9 \times 10^{-7}$. 

\begin{table}
 \begin{center}
  \caption{Mean value and standard deviation of each abundance ratio for the
low-$\alpha$ group and the high-$\alpha$ group. Y and Ba are taken from
NS11. The $p$-value between the two groups is also given.
 \label{tab:med_std}}
  \vspace{1mm}
   \begin{tabular}{lrrcrrcc}
\hline \hline
& \multicolumn{2}{c}{low-$\alpha$} 	& &  \multicolumn{2}{c}{high-$\alpha$} & \\
\multicolumn{1}{c}{} &  \multicolumn{1}{c}{\emph{\~{x}}} & \multicolumn{1}{c}{$\sigma$} & &  \multicolumn{1}{c}{\emph{\~{x}}} & \multicolumn{1}{c}{$\sigma$} & & $p$-value \\
\hline
{[Sc/Fe]} & $-$0.107 & 0.062 & & 0.067 & 0.063 & & 0.000   \\
{[Zr/Fe]} & $-$0.058 & 0.061 & & 0.069 & 0.131 & & 0.017   \\
{[La/Fe]} & $-$0.022 & 0.065 & & 0.004 & 0.076 & & 0.589   \\
{[Ce/Fe]} & $-$0.012 & 0.087 & & $-$0.058 & 0.114 & &  0.340   \\
{[Nd/Fe]} & 0.217 & 0.137 && 0.299 & 0.098 & &0.614   \\
{[Eu/Fe]} & 0.222 & 0.097 && 0.161 & 0.068 & &0.173   \\
{[La/Zr]} & 0.037 & 0.061 & & $-$0.066 & 0.084 & &0.005   \\
{[$ls$/Fe]} & $-$0.109 & 0.057& & 0.049 & 0.117 & &0.004   \\
{[$hs$/Fe]} & 0.005 & 0.056 & & 0.038 & 0.084 & &0.303   \\
{[$hs$/$ls$]} & 0.114 & 0.046& & $-$0.010 & 0.065 & &0.000   \\
{[Y/Eu]} & $-$0.382 & 0.101& & $-$0.133 & 0.122 & &0.000   \\
{[$ls$/Eu]} & $-$0.332 & 0.092 && $-$0.112 & 0.126 & &0.001   \\
{[Ba/Eu]} & $-$0.387 & 0.122 & &  $-$0.253 &  0.122 & &  0.024 \\
{[La/Eu]} & $-$0.244 & 0.064 && $-$0.157 & 0.082 && 0.173   \\
{[$hs$/Eu]} & $-$0.218 & 0.082 && $-$0.122 & 0.081 && 0.022   \\
{[Y/Fe]} & $-$0.160 & 0.064& & 0.028 & 0.107 & &0.000   \\
{[Ba/Fe]} & $-$0.165 & 0.070&& $-$0.092 & 0.115 & &0.303   \\
{[Ba/Y]} & $-$0.005 & 0.055 && $-$0.120 & 0.030 && 0.000   \\
\hline \hline
  \end{tabular} 
\\
 \end{center}
\end{table}

For abundances relative to Fe (e.g., [Zr/Fe]), Sc and Zr show a separation 
between the low- and high-$\alpha$ groups. For both elements, the
high-$\alpha$ group has a higher [X/Fe] ratio than the low-$\alpha$ group. 

We find that the separation between the low- and high-$\alpha$ groups is
detected for [Y/Fe] despite having a smaller sample size than NS11. 
In contrast to
the first $s$-process peak elements Y and Zr, the abundance ratios for elements
belonging to the second peak [La/Fe], [Ce/Fe], and [Nd/Fe] show no significant
difference between the low- and high-$\alpha$ groups. 
The second $s$-process peak element Ba, measured by NS11, also shows no
separation in [Ba/Fe] between the two groups. 

\begin{figure}
\begin{center}
\includegraphics[]{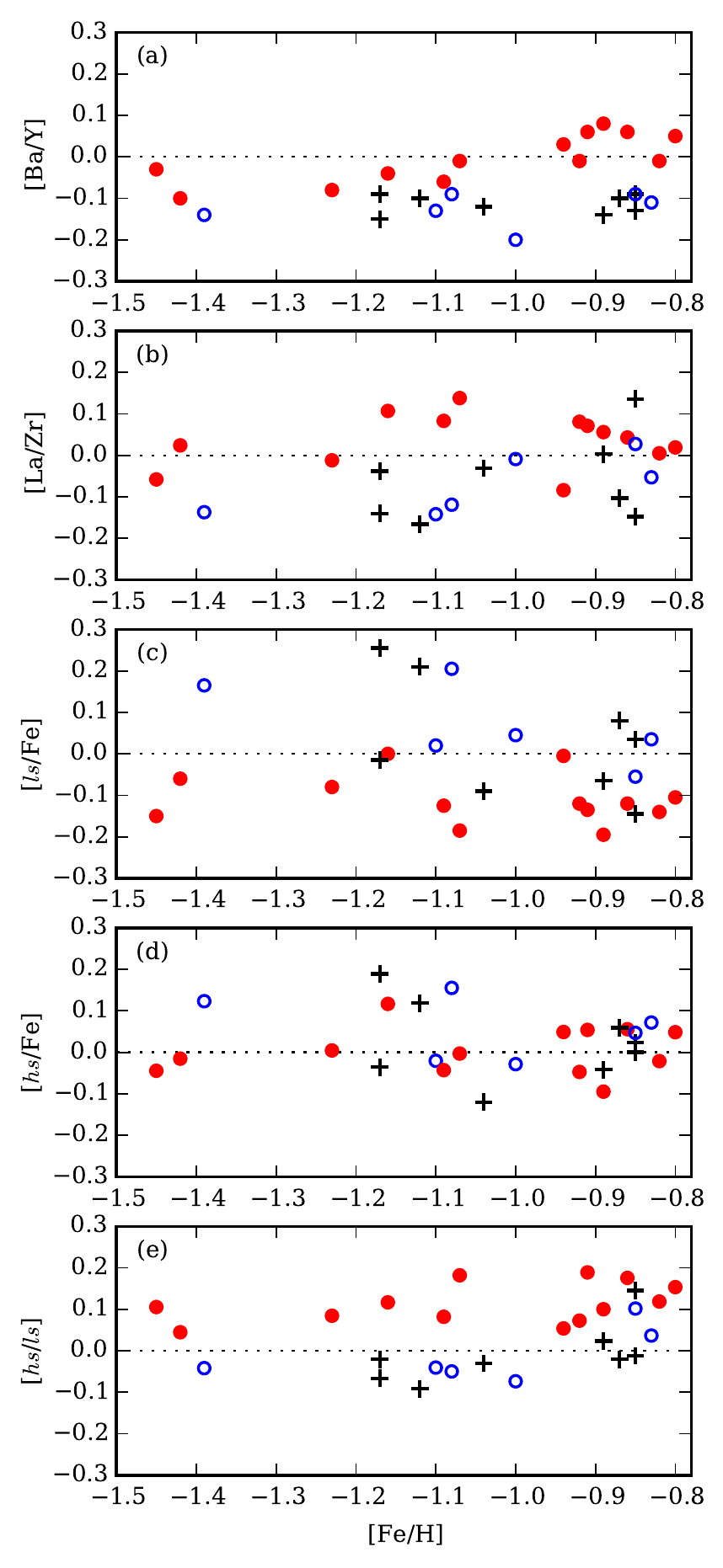} 
\caption{[Ba/Y] (from NS11), [La/Zr], [$ls$/Fe], [$hs$/Fe], and [$hs$/$ls$]
versus [Fe/H] where $ls$ is the average abundance of Y and Zr and $hs$ is the
average abundance of Ba, La, Nd, and Ce. The red circles indicate low-$\alpha$
stars, the blue open circles indicate high-$\alpha$ stars, and the plus signs
indicate TD stars.} 
\label{fig:sprocess}
\end{center}
\end{figure}

We confirm the separation in [Ba/Y] reported by NS11 
(see Figure~\ref{fig:sprocess}(a)). 
In Figure~\ref{fig:sprocess}(b) we plot 
[La/Zr] and are able to reproduce the separation between a light-$s$ ($ls$)
element at the first $s$-process peak and a heavy-$s$ ($hs$) element at the
second peak. 
The $p$-value of 0.005 is lower than the individual
$p$-values for [Zr/Fe] (0.017) and [La/Fe] (0.589) demonstrating
that a clear separation between the low- and high-$\alpha$ populations can be
seen when considering certain combinations of elements. 

Assuming that the higher [La/Zr] ratios in the low-$\alpha$ stars are due to
enrichment from AGB stars, we use the model predictions of
\citet{Fishlock:2014aa} with [Fe/H] = $-1.2$ to estimate the mass range of
those AGB stars. Starting at lower masses (1 to 3~$\Msun$), calculations
predict [La/Zr] ratios above solar; for example, the 3~$\Msun$ model predicts a
final [La/Zr] ratio of 0.32 dex. Moving to the intermediate masses (3.25 to
7~$\Msun$), calculations predict [La/Zr] ratios below solar; for example, the
6~$\Msun$ model predicts a final [La/Zr] ratio of $-0.68$ dex. Therefore, based
on the [La/Zr] ratio, AGBs stars in the range 1 to 3~$\Msun$ could be
responsible for the abundance difference between the low- and high-$\alpha$
groups. 

For stars in the range $-1.5 <$ [Fe/H] $< -1$, there is a possible separation
in [Eu/Fe] where the low-$\alpha$ group has a higher [Eu/Fe] abundance compared
to the high-$\alpha$ group. However, higher metallicity stars show no
difference between the two groups and the hint of a separation is not
statistically significant with a $p$-value of 0.173. We return to this
intriguing possibility of a separation in Section 5. 

The $s$-process indicators are the average abundance of the elements at each of
the two $s$-process peaks. We include the Y and Ba results from NS11 and define
[$ls$/Fe] and [$hs$/Fe]\footnote{As noted, the Solar System abundance of Nd has
a $\sim$40\% contribution from the $r$-process. Had we excluded Nd from $hs$,
as did \citet{Lugaro:2012aa}, our conclusions would be unchanged.} to be: 

\begin{equation}
[ls/{\rm Fe}] = ([{\rm Y}/{\rm Fe}]+[{\rm Zr}/{\rm Fe}])/2
\end{equation}

and

\begin{equation}
[hs/{\rm Fe}] = ([{\rm Ba}/{\rm Fe}]+[{\rm La}/{\rm Fe}]+[{\rm Ce}/{\rm Fe}]+[{\rm Nd}/{\rm Fe}])/4.
\end{equation}

The [$ls$/Fe] and [$hs$/Fe] ratios against [Fe/H] are presented in
Figure~\ref{fig:sprocess}(c) and~\ref{fig:sprocess}(d).  The low-$\alpha$ stars
have a lower [$ls$/Fe] ratio than the high-$\alpha$ stars. 
For [$hs$/Fe], there is no apparent difference between the two
populations. 

Figure~\ref{fig:sprocess}(e) shows there is a clear separation in the
$s$-process indicator [$hs$/$ls$]  between the low- and high-$\alpha$ groups.
The low-$\alpha$ population has a higher [$hs$/$ls$] ratio than the
high-$\alpha$ population with a mean [$hs$/$ls$] value of $0.11$ dex compared
to $-0.01$ dex for the high-$\alpha$ population. 
All the low-$\alpha$ stars have a
supersolar [$hs$/$ls$] ratio and this would be expected if the stars were
enriched by previous generations of low-mass AGB stars. For example, the
low-mass models of \citet{Fishlock:2014aa} end the AGB phase with [$hs$/$ls$]
ratios of between 0.45 and 0.59 dex, excluding the 1~$\Msun$ model which
experiences minimal third dredge-up. In contrast, the intermediate-mass models
have final [$hs$/$ls$] ratios between $-0.71$ and $-0.41$ dex. The differences
between the low-mass and intermediate-mass models are a result of the different
neutron sources \citep{Lugaro:2012aa}. 

\begin{figure}
\begin{center}
\includegraphics[]{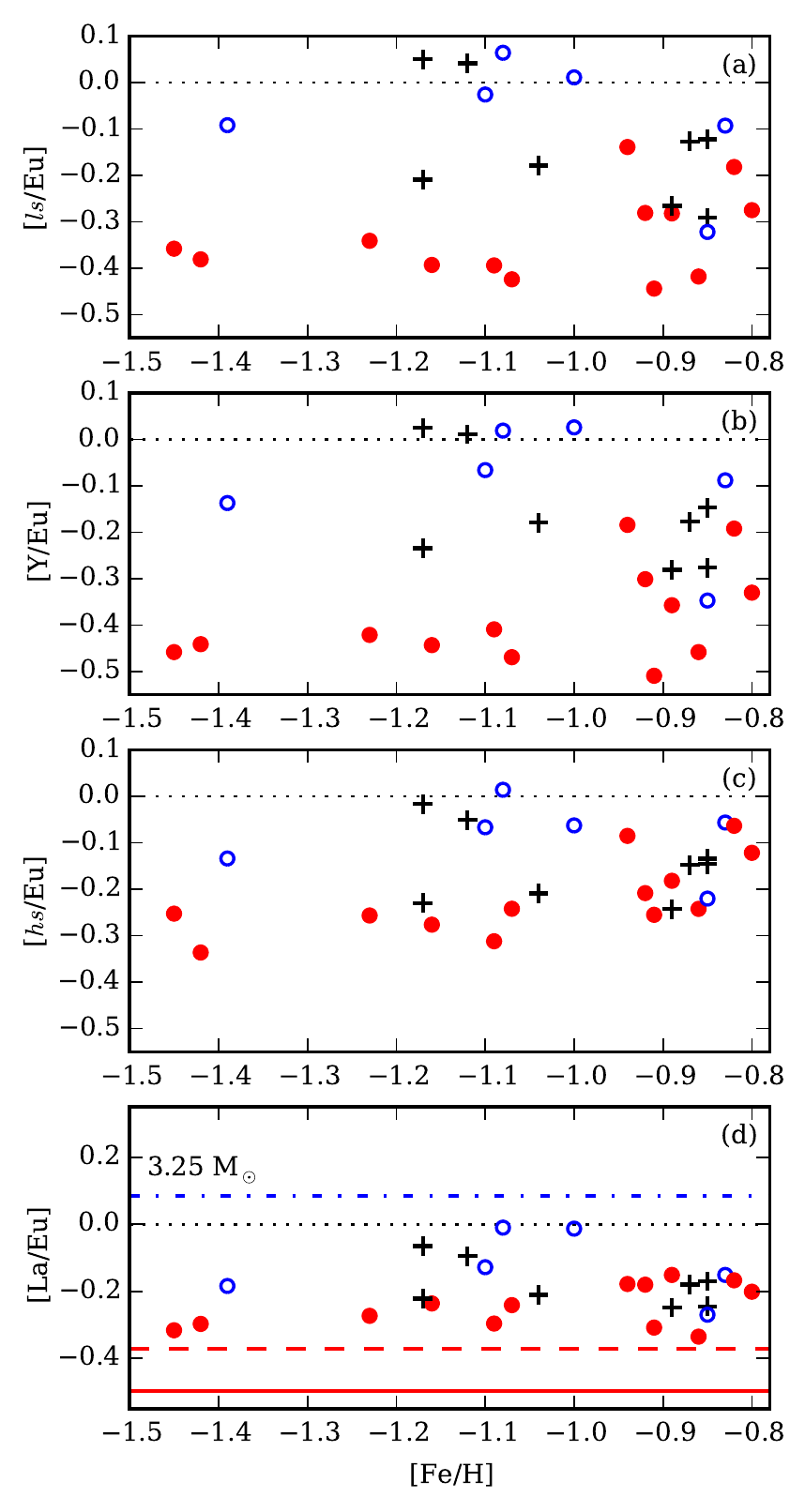} 
\caption{[$ls$/Eu], [Y/Eu], [$hs$/Eu], and [La/Eu] versus [Fe/H] where $ls$ is
the average abundance of Y and Zr and $hs$ is the average abundance of Ba, La,
Nd, and Ce. The red circles indicate low-$\alpha$ stars, the blue open circles
indicate high-$\alpha$ stars, and the plus signs indicate TD stars. The red
solid line represents the $r$-process ratio for [La/Eu]  and the red dashed
line its upper uncertainty \citep{Winckler:2006aa}. The final surface
abundance of [La/Eu] for the 3.25~$\Msun$, $Z~=~0.001$ AGB model from
\citet{Fishlock:2014aa} is also indicated (dash-dotted line).} 
\label{fig:sprocess2}
\end{center}
\end{figure}

The abundance ratio of $s$-process elements to $r$-process elements offers a
powerful diagnostic for Galactic chemical evolution
\citep{Wheeler:1989aa,McWilliam:1997aa}. In Figure~\ref{fig:sprocess2} we
present the abundance ratios of [$ls$/Eu], [Y/Eu], [$hs$/Eu], and  [La/Eu]
against [Fe/H]. 
The low-$\alpha$ stars have a lower
[$ls$/Eu] ratio compared to the high-$\alpha$ stars with the separation more
evident at [Fe/H] $\lesssim -1$. There is also an apparent increase in
[$ls$/Eu] with increasing [Fe/H] for the low-$\alpha$ group which is not seen
in the high-$\alpha$ group.

Figure~\ref{fig:sprocess2}(b) presents the abundance ratio of the $ls$
element, Y, relative to Eu. The low-$\alpha$ group has a lower abundance of
[Y/Eu] compared to the high-$\alpha$ group with a $p$-value of $2.2 \times
10^{-4}$. This $p$-value is more significant than the [$ls$/Eu] separation. We
also find that the [Y/Eu] abundance for the low-$\alpha$ stars with [Fe/H] $<$
$-1$ is almost constant at approximately $-0.4$ dex with minimal scatter.
Similar behaviour is also seen for [$ls$/Fe]. 
 
The separation in [$hs$/Eu] is not as clear as that of [$ls$/Eu], although
it has a $p$-value of 0.022, just below the chosen significance level. The
low-$\alpha$ group has a [$hs$/Eu] ratio that is closer to the $r$-process
contribution of the $hs$ elements and Eu in the Solar System than the
high-$\alpha$ group.  As with [$ls$/Fe], there is an apparent increase in
[$hs$/Eu] with increasing [Fe/H] for the low-$\alpha$ stars. 

Figure~\ref{fig:sprocess2}(d) presents the abundance ratios of the $hs$
element, La, relative to Eu where all the stars have a subsolar [La/Eu] ratio.
Unlike the separation found for [$hs$/Eu], we do not find a separation in
[La/Eu] between the two groups. We also show the $r$-process contribution to
the Solar System abundances of La and Eu (and its uncertainty) in
Figure~\ref{fig:sprocess2}(d) as determined by \citet{Winckler:2006aa}. The
theoretical prediction of an AGB stellar model with an initial mass of
3.25~$\Msun$ at a metallicity of [Fe/H] = $-1.2$, from
\citet{Fishlock:2014aa} is  also presented in Figure~\ref{fig:sprocess2}.
Of all the models calculated by \citet{Fishlock:2014aa}, the lowest [La/Eu]
ratio of 0.08 dex occurs for the 3.25 M$_{\odot}$ model. 

\newpage

\section{Discussion}

We have studied chemical abundances in a subsample
of low- and high-$\alpha$ stars from NS10 in which the 
low-$\alpha$ stars were likely accreted from satellite dwarf galaxies
whereas the high-$\alpha$ stars are representative of Galactic halo objects.
We now seek to understand the abundance differences between these
populations. 

For Sc, we confirm and extend upon results from \citet{Nissen:2000ab} in which
the low-$\alpha$ stars have a lower [Sc/Fe] ratio than the high-$\alpha$ stars.
That is, Sc appears to behave like an $\alpha$-capture element. The simplest
explanation is that Sc is preferentially made in higher mass stars along with
the $\alpha$ elements and that Type Ia SNe have a negligible contribution to Sc 
\citep{Iwamoto:1999aa,Woosley:1995aa,Romano:2010aa}. Indeed,
\citet{Nissen:2016aa} find that the [Sc/Fe] ratio exhibits a similar behaviour
with age as [Mg/Fe] in their sample of solar twins. The low-$\alpha$ stars have
low [Sc/Fe] ratios due to the extra contribution of Fe from Type Ia SNe. 

As noted in the introduction, there is another chemical signature 
besides [$\alpha$/Fe]\footnote{We note that there is some overlap in
[$\alpha$/Fe] between dSph stars and Galactic halo stars with extreme
retrograde orbits \citep{Fulbright:2002aa,Stephens:2002aa}. Additionally, the
most metal-poor stars in dSph appear to have halo-like [$\alpha$/Fe] ratios
\citep{Frebel:2010aa,Norris:2010ac,Venn:2012aa}.} that distinguishes halo
stars from their dSph counterparts, namely, the neutron-capture
element abundance ratios. In dSph, roughly two-thirds of the stars show a
higher [Ba/Y] ratio and a lower [Y/Eu] ratio compared to Galactic halo stars
\citep{Venn:2004aa,Tolstoy:2009aa}. 
\citet{McWilliam:2013aa} analysed three Sagittarius dSph stars and found
low [$\alpha$/Fe], [Na/Fe], [Al/Fe], and [Cu/Fe] ratios along with a
high [La/Y] ratio relative to the Galactic halo. The study of the Carina dSph
by \citet{Venn:2012aa} also found that the low-metallicity stars have slightly
lower [Y/Eu] ratios (and slightly higher [Ba/Y] ratios) when compared to
Galactic stars. The dSph of today are chemically different to the galaxies that
merged early during the formation of the Galaxy. 

The NS11 study found that the low-$\alpha$ stars show a higher [Ba/Y]
ratio than the high-$\alpha$ stars. We also find differences for [Zr/Fe],
[La/Zr], [$ls$/Fe], [$hs$/$ls$], [Y/Eu], [Ba/Eu], [$ls$/Eu], and [$hs$/Eu]
between the two populations. These findings reveal that the low-$\alpha$
population displays many chemical signatures also seen in dSph thereby
supporting the accretion hypothesis from NS10. 

We now turn our attention to the possible sources that could produce these
abundance differences among the neutron-capture elements. 
Due to the slower rate of chemical evolution, it has been suggested that the
dSph and low-$\alpha$ population have been enriched from low-metallicity,
low-mass AGB stars (NS11, \citealt{Venn:2012aa}). 
Intermediate-mass AGB stars ($> 3$~$\Msun$) produce subsolar
[Ba/Y] ratios \citep[e.g.,][]{Karakas:2014ab}, i.e., these objects cannot explain 
the behaviour of [Ba/Y] in the low-$\alpha$ population. On the other hand,
low-metallicity, low-mass AGB stars ($\lesssim 3$~$\Msun$) produce [Ba/Y]
ratios above solar \citep[e.g.,][]{Cristallo:2011aa,Fishlock:2014aa}, and this is
consistent with the low-$\alpha$ population. The high [La/Zr] and [$hs$/$ls$]
ratios for the low-$\alpha$ population further support the hypothesis of 
enrichment from low-metallicity, low-mass AGBs in the low-$\alpha$ populations
as a result of the slower rate of chemical evolution (e.g., in low-mass AGBs,
the first $s$-process peak elements are bypassed in favour of second
$s$-process peak elements). 

One issue with this scenario is that we would therefore expect the abundance of
Y to be greater than (or equal to) Zr, i.e., [Y/Fe] $\ge$ [Zr/Fe] (e.g.,
\citealt{Bisterzo:2014aa,Trippella:2016aa}). The observations indicate
otherwise, and this issue has also been identified in CEMP-s stars
\citep{Abate:2015aa}. The reason for the discrepancy between observations and
theory is unknown. As noted in Section 3.1, while the abundances of Y and Zr
follow each other, there may be a zero-point offset due to errors in the atomic
data (e.g., $gf$ values). 

If low-metallicity, low-mass stars have contributed to chemical enrichment of
the low-$\alpha$ population, then we can place constraints on the enrichment
timescales involved in the earliest dwarf galaxies. Models from
\citet{Fishlock:2014aa} predict that low-metallicity, low-mass ($\lesssim
3$~$\Msun$) stars have lifetimes between $0.29 \times 10^9$ yr to $6.8 \times
10^9$ yr. 

We also present measurements of Eu; the $r$-process is responsible
for producing the majority of Eu in the Solar System \citep{Sneden:2008aa}.
The evolution of [Eu/Fe] for both Galactic stars and dSph stars generally
follows [$\alpha$/Fe] \citep{Venn:2004aa,Tolstoy:2009aa} indicating a common
origin between $\alpha$ elements and Eu, most likely to be massive stars. If
true, then we would expect the low-$\alpha$ stars to have a lower [Eu/Fe]
abundance than the high-$\alpha$ stars at a given metallicity. 
However, there is some evidence that the low-$\alpha$ stars have a higher 
[Eu/Fe] ratio than the high-$\alpha$ stars, in contrast to our expectations.
This separation is only observed at [Fe/H] $<$ $-1$ and includes half the
sample. When using the whole metallicity range, the KS test does not find a
statistically significant separation between the two populations (see
Table~\ref{tab:med_std}). 
 
If the low-$\alpha$ stars do have a higher Eu abundance than the high-$\alpha$
stars at lower metallicities, it would suggest that the $r$-process contributed
more to the chemical evolution of the low-$\alpha$ stars at a given
metallicity. This is difficult to reconcile with chemical evolution models
which support the idea that intense galactic winds remove Eu from a system
\citep{Lanfranchi:2008aa}. Another possibility is that the $\alpha$ and Eu
abundances are due to different initial mass functions rather than the addition
of SNe Ia material \citep{Ishimaru:1999aa,Travaglio:1999aa,Kobayashi:2006aa}.
To determine if there is the possibility of a statistically significant
separation, additional observations of Eu for the NS10 sample are required,
particularly for stars with [Fe/H] $<$ $-1$. 

We observe constant [Y/Eu] abundances ($\sim -0.4$ dex) in the lower
metallicity low-$\alpha$ stars ($-1.5 \leq$ [Fe/H] $\leq -1$) suggesting that Y
and Eu have a common origin or timescale for enrichment, most likely to be the
$r$-process, at these metallicities. The dSph chemical evolution models of
\citet{Lanfranchi:2008aa} show constant [Y/Eu] ratios  up to a [Fe/H] ratio of
approximately $-1.5$. The ratio of [Y/Eu] then increases at higher 
metallicities due to the production of Y by AGB stars. 

We observe lower [Y/Eu], [$ls$/Eu], and [$hs$/Eu] ratios in the
low-$\alpha$ stars compared to the high-$\alpha$ stars with the separation
becoming more noticeable at lower metallicities. The study of the Carina dSph
by \citet{Venn:2012aa} also found that the low-metallicity stars have slightly
lower [Y/Eu] ratios (and slightly higher [Ba/Y] ratios) when compared to
Galactic stars. \citet{Venn:2012aa} concluded that these differences are a
result of metal-poor AGB stars contributing to the chemical enrichment of
Carina, and this may also apply to the low-$\alpha$ population. 

\section{Conclusions}

We found that the low-$\alpha$ and high-$\alpha$ populations of NS10 separate
when using the abundance ratios of [Sc/Fe], [Zr/Fe], [La/Zr], [$ls$/Fe],
[$hs$/$ls$], [Y/Eu], [Ba/Eu], [$ls$/Eu], and [$hs$/Eu]. These differences in
chemical abundance ratios have only been detected using a differential analysis
relative to a reference star with comparable stellar parameters to minimise the
errors in abundance measurements. The separation observed in [La/Zr] between
the low- and high-$\alpha$ groups confirms the results of NS11 who find a
separation in [Ba/Y], where Zr and Y are first $s$-process peak elements and La
and Ba are second $s$-process peak elements. 

The low abundance of [Y/Eu] found in the low-$\alpha$ population compared to
the high-$\alpha$ population matches one of the chemical signatures of
present-day dSph. \citet{Venn:2004aa} discard the possibility that the Galactic
halo could consist of low-mass dwarf galaxies that have continuously merged
with the Galaxy as high [Ba/Y] and low [Y/Eu] ratios are not observed in
Galactic halo stars. 
The idea of more massive mergers
with the Galaxy was not ruled out by \citet{Venn:2004aa}.  In addition, models
of AGB stars are unable to explain the lower [$hs$/Eu] ratios observed in the
low-$\alpha$ stars compared to the high-$\alpha$ stars. 

The results presented here offer new and important observational evidence
regarding the nature of the Galactic halo. It is apparent from the study of
NS10 and NS11 that there are at least two populations in the Galactic halo, one
that formed in-situ as well as one accreted from dwarf galaxies. The low [Y/Eu]
abundances in dSph have been attributed to the contribution of metal-poor AGB
stars \citep{Venn:2004aa}. Low-metallicity AGB stars could also be responsible
for the high [Ba/Y] abundances.  Only with detailed chemical evolution models,
along with measurements of additional $r$-process elements beyond Eu, will
there be a better understanding of the chemical enrichment of the earliest
dwarf galaxies that merged with the Galactic halo. In particular, through a
better understanding of the $s$-process, we will be able to provide constraints
on the formation site(s) of the $r$-process. 

\section{Acknowledgements}

We thank the referee, Maurizio Busso, for helpful comments. 
This work has been supported by the Australian Research Council (grants
FT110100475 and FT140100554) and the European Union FP7 programme through ERC
grant number 320360. 
JM thanks support by FAPESP (2012/24392-2 and 2014/18100-4). 
Funding for the Stellar Astrophysics Centre is provided by The Danish National
Research Foundation (Grant DNRF106). 
Australian access to the Magellan Telescopes was supported through the National
Collaborative Research Infrastructure Strategy of the Australian Federal
Government.

\label{lastpage}


\begin{thebibliography}{64}
\expandafter\ifx\csname natexlab\endcsname\relax\def\natexlab#1{#1}\fi

\bibitem[{{Abate} {et~al}\mbox{.}(2015){Abate}, {Pols}, {Izzard}, \&
  {Karakas}}]{Abate:2015aa}
{Abate} C., {Pols} O.~R., {Izzard} R.~G., {Karakas} A.~I., 2015, \aap, 581, A22

\bibitem[{{Asplund} {et~al}\mbox{.}(2009){Asplund}, {Grevesse}, {Sauval}, \&
  {Scott}}]{Asplund:2009aa}
{Asplund} M., {Grevesse} N., {Sauval} A.~J., {Scott} P., 2009, \araa, 47, 481

\bibitem[{{Belokurov} {et~al}\mbox{.}(2006){Belokurov}, {Zucker}, {Evans},
  {Gilmore}, {Vidrih}, {Bramich}, {Newberg}, {Wyse}, {Irwin}, {Fellhauer},
  {Hewett}, {Walton}, {Wilkinson}, {Cole}, {Yanny}, {Rockosi}, {Beers}, {Bell},
  {Brinkmann}, {Ivezi{\'c}}, \& {Lupton}}]{Belokurov:2006aa}
{Belokurov} V. {et~al.}, 2006, \apjl, 642, L137

\bibitem[{{Bernstein} {et~al}\mbox{.}(2003){Bernstein}, {Shectman}, {Gunnels},
  {Mochnacki}, \& {Athey}}]{Bernstein:2003aa}
{Bernstein} R., {Shectman} S.~A., {Gunnels} S.~M., {Mochnacki} S., {Athey}
  A.~E., 2003, in \procspie, Vol. 4841, Instrument Design and Performance for
  Optical/Infrared Ground-based Telescopes, {Iye} M., {Moorwood} A.~F.~M.,
  eds., pp. 1694--1704

\bibitem[{{Bisterzo} {et~al}\mbox{.}(2014){Bisterzo}, {Travaglio}, {Gallino},
  {Wiescher}, \& {K{\"a}ppeler}}]{Bisterzo:2014aa}
{Bisterzo} S., {Travaglio} C., {Gallino} R., {Wiescher} M., {K{\"a}ppeler} F.,
  2014, \apj, 787, 10

\bibitem[{{Bullock} \& {Johnston}(2005)}]{Bullock:2005aa}
{Bullock} J.~S., {Johnston} K.~V., 2005, \apj, 635, 931

\bibitem[{{Busso} {et~al}\mbox{.}(2001){Busso}, {Gallino}, {Lambert},
  {Travaglio}, \& {Smith}}]{Busso:2001aa}
{Busso} M., {Gallino} R., {Lambert} D.~L., {Travaglio} C., {Smith} V.~V., 2001,
  \apj, 557, 802

\bibitem[{{Busso}, {Gallino} \& {Wasserburg}(1999){Busso}, {Gallino}, \&
  {Wasserburg}}]{Busso:1999aa}
{Busso} M., {Gallino} R., {Wasserburg} G.~J., 1999, \araa, 37, 239

\bibitem[{{Carney} {et~al}\mbox{.}(1997){Carney}, {Wright}, {Sneden}, {Laird},
  {Aguilar}, \& {Latham}}]{Carney:1997aa}
{Carney} B.~W., {Wright} J.~S., {Sneden} C., {Laird} J.~B., {Aguilar} L.~A.,
  {Latham} D.~W., 1997, \aj, 114, 363

\bibitem[{{Casey}(2014)}]{Casey:2014aa}
{Casey} A.~R., 2014, PhD thesis, Australian National University

\bibitem[{{Casey} {et~al}\mbox{.}(2014){Casey}, {Keller}, {Alves-Brito},
  {Frebel}, {Da Costa}, {Karakas}, {Yong}, {Schlaufman}, {Jacobson}, {Yu}, \&
  {Fishlock}}]{Casey:2014ab}
{Casey} A.~R. {et~al.}, 2014, \mnras, 443, 828

\bibitem[{{Castelli} \& {Kurucz}(2003)}]{Castelli:2003aa}
{Castelli} F., {Kurucz} R.~L., 2003, in IAU Symposium, Vol. 210, Modelling of
  Stellar Atmospheres, {Piskunov} N., {Weiss} W.~W., {Gray} D.~F., eds., p. 20P

\bibitem[{{Cristallo} {et~al}\mbox{.}(2011){Cristallo}, {Piersanti},
  {Straniero}, {Gallino}, {Dom{\'{\i}}nguez}, {Abia}, {Di Rico}, {Quintini}, \&
  {Bisterzo}}]{Cristallo:2011aa}
{Cristallo} S. {et~al.}, 2011, \apjs, 197, 17

\bibitem[{{Eggen}, {Lynden-Bell} \& {Sandage}(1962){Eggen}, {Lynden-Bell}, \&
  {Sandage}}]{Eggen:1962aa}
{Eggen} O.~J., {Lynden-Bell} D., {Sandage} A.~R., 1962, \apj, 136, 748

\bibitem[{{Fishlock} {et~al}\mbox{.}(2014){Fishlock}, {Karakas}, {Lugaro}, \&
  {Yong}}]{Fishlock:2014aa}
{Fishlock} C.~K., {Karakas} A.~I., {Lugaro} M., {Yong} D., 2014, \apj, 797, 44

\bibitem[{{Frebel} \& {Norris}(2015)}]{Frebel:2015aa}
{Frebel} A., {Norris} J.~E., 2015, \araa, 53, 631

\bibitem[{{Frebel} {et~al}\mbox{.}(2010){Frebel}, {Simon}, {Geha}, \&
  {Willman}}]{Frebel:2010aa}
{Frebel} A., {Simon} J.~D., {Geha} M., {Willman} B., 2010, \apj, 708, 560

\bibitem[{{Fulbright}(2002)}]{Fulbright:2002aa}
{Fulbright} J.~P., 2002, \aj, 123, 404

\bibitem[{{Herwig}(2005)}]{Herwig:2005aa}
{Herwig} F., 2005, \araa, 43, 435

\bibitem[{{Ibata}, {Gilmore} \& {Irwin}(1994){Ibata}, {Gilmore}, \&
  {Irwin}}]{Ibata:1994aa}
{Ibata} R.~A., {Gilmore} G., {Irwin} M.~J., 1994, \nat, 370, 194

\bibitem[{{Ishimaru} \& {Wanajo}(1999)}]{Ishimaru:1999aa}
{Ishimaru} Y., {Wanajo} S., 1999, \apjl, 511, L33

\bibitem[{{Ivans} {et~al}\mbox{.}(2003){Ivans}, {Sneden}, {James}, {Preston},
  {Fulbright}, {H{\"o}flich}, {Carney}, \& {Wheeler}}]{Ivans:2003aa}
{Ivans} I.~I., {Sneden} C., {James} C.~R., {Preston} G.~W., {Fulbright} J.~P.,
  {H{\"o}flich} P.~A., {Carney} B.~W., {Wheeler} J.~C., 2003, \apj, 592, 906

\bibitem[{{Iwamoto} {et~al}\mbox{.}(1999){Iwamoto}, {Brachwitz}, {Nomoto},
  {Kishimoto}, {Umeda}, {Hix}, \& {Thielemann}}]{Iwamoto:1999aa}
{Iwamoto} K., {Brachwitz} F., {Nomoto} K., {Kishimoto} N., {Umeda} H., {Hix}
  W.~R., {Thielemann} F.-K., 1999, \apjs, 125, 439

\bibitem[{{Karakas} \& {Lattanzio}(2014)}]{Karakas:2014ab}
{Karakas} A.~I., {Lattanzio} J.~C., 2014, \pasa, 31, e030

\bibitem[{{King}(1997)}]{King:1997aa}
{King} J.~R., 1997, \aj, 113, 2302

\bibitem[{{Kobayashi} {et~al}\mbox{.}(2006){Kobayashi}, {Umeda}, {Nomoto},
  {Tominaga}, \& {Ohkubo}}]{Kobayashi:2006aa}
{Kobayashi} C., {Umeda} H., {Nomoto} K., {Tominaga} N., {Ohkubo} T., 2006,
  \apj, 653, 1145

\bibitem[{{Lanfranchi}, {Matteucci} \& {Cescutti}(2008){Lanfranchi},
  {Matteucci}, \& {Cescutti}}]{Lanfranchi:2008aa}
{Lanfranchi} G.~A., {Matteucci} F., {Cescutti} G., 2008, \aap, 481, 635

\bibitem[{{Lawler}, {Bonvallet} \& {Sneden}(2001){Lawler}, {Bonvallet}, \&
  {Sneden}}]{Lawler:2001aa}
{Lawler} J.~E., {Bonvallet} G., {Sneden} C., 2001, \apj, 556, 452

\bibitem[{{Lawler} {et~al}\mbox{.}(2001){Lawler}, {Wickliffe}, {den Hartog}, \&
  {Sneden}}]{Lawler:2001ab}
{Lawler} J.~E., {Wickliffe} M.~E., {den Hartog} E.~A., {Sneden} C., 2001, \apj,
  563, 1075

\bibitem[{{Lodders}(2003)}]{Lodders:2003aa}
{Lodders} K., 2003, \apj, 591, 1220

\bibitem[{{Lugaro} {et~al}\mbox{.}(2012){Lugaro}, {Karakas}, {Stancliffe}, \&
  {Rijs}}]{Lugaro:2012aa}
{Lugaro} M., {Karakas} A.~I., {Stancliffe} R.~J., {Rijs} C., 2012, \apj, 747, 2

\bibitem[{{Matteucci} \& {Greggio}(1986)}]{Matteucci:1986aa}
{Matteucci} F., {Greggio} L., 1986, \aap, 154, 279

\bibitem[{{McWilliam}(1997)}]{McWilliam:1997aa}
{McWilliam} A., 1997, \araa, 35, 503

\bibitem[{{McWilliam}, {Wallerstein} \& {Mottini}(2013){McWilliam},
  {Wallerstein}, \& {Mottini}}]{McWilliam:2013aa}
{McWilliam} A., {Wallerstein} G., {Mottini} M., 2013, \apj, 778, 149

\bibitem[{{Nissen}(2013)}]{Nissen:2013aa}
{Nissen} P.~E., 2013, {Chemical Abundances as Population Tracers in Planets,
  Stars and Stellar Systems: Volume 5}, {Oswalt} T.~D., {Gilmore} G., eds.,
  Springer Netherlands, Dordrecht, pp. 21--54

\bibitem[{{Nissen}(2016)}]{Nissen:2016aa}
{Nissen} P.~E., 2016, \aap, 593, A65

\bibitem[{{Nissen} {et~al}\mbox{.}(2000){Nissen}, {Chen}, {Schuster}, \&
  {Zhao}}]{Nissen:2000ab}
{Nissen} P.~E., {Chen} Y.~Q., {Schuster} W.~J., {Zhao} G., 2000, \aap, 353, 722

\bibitem[{{Nissen} \& {Schuster}(2010)}]{Nissen:2010aa}
{Nissen} P.~E., {Schuster} W.~J., 2010, \aap, 511, L10

\bibitem[{{Nissen} \& {Schuster}(2011)}]{Nissen:2011aa}
{Nissen} P.~E., {Schuster} W.~J., 2011, \aap, 530, A15

\bibitem[{{Nomoto}, {Kobayashi} \& {Tominaga}(2013){Nomoto}, {Kobayashi}, \&
  {Tominaga}}]{Nomoto:2013aa}
{Nomoto} K., {Kobayashi} C., {Tominaga} N., 2013, \araa, 51, 457

\bibitem[{{Norris} {et~al}\mbox{.}(2010){Norris}, {Yong}, {Gilmore}, \&
  {Wyse}}]{Norris:2010ac}
{Norris} J.~E., {Yong} D., {Gilmore} G., {Wyse} R.~F.~G., 2010, \apj, 711, 350

\bibitem[{{Pillepich} {et~al}\mbox{.}(2014){Pillepich}, {Vogelsberger},
  {Deason}, {Rodriguez-Gomez}, {Genel}, {Nelson}, {Torrey}, {Sales},
  {Marinacci}, {Springel}, {Sijacki}, \& {Hernquist}}]{Pillepich:2014aa}
{Pillepich} A. {et~al.}, 2014, \mnras, 444, 237

\bibitem[{{Preston} \& {Sneden}(2000)}]{Preston:2000aa}
{Preston} G.~W., {Sneden} C., 2000, \aj, 120, 1014

\bibitem[{{Ram{\'{\i}}rez}, {Mel{\'e}ndez} \&
  {Chanam{\'e}}(2012){Ram{\'{\i}}rez}, {Mel{\'e}ndez}, \&
  {Chanam{\'e}}}]{Ramirez:2012aa}
{Ram{\'{\i}}rez} I., {Mel{\'e}ndez} J., {Chanam{\'e}} J., 2012, \apj, 757, 164

\bibitem[{{Romano} {et~al}\mbox{.}(2010){Romano}, {Karakas}, {Tosi}, \&
  {Matteucci}}]{Romano:2010aa}
{Romano} D., {Karakas} A.~I., {Tosi} M., {Matteucci} F., 2010, \aap, 522, A32

\bibitem[{{Searle} \& {Zinn}(1978)}]{Searle:1978aa}
{Searle} L., {Zinn} R., 1978, \apj, 225, 357

\bibitem[{{Shetrone}, {C{\^o}t{\'e}} \& {Sargent}(2001){Shetrone},
  {C{\^o}t{\'e}}, \& {Sargent}}]{Shetrone:2001aa}
{Shetrone} M.~D., {C{\^o}t{\'e}} P., {Sargent} W.~L.~W., 2001, \apj, 548, 592

\bibitem[{{Simmerer} {et~al}\mbox{.}(2004){Simmerer}, {Sneden}, {Cowan},
  {Collier}, {Woolf}, \& {Lawler}}]{Simmerer:2004aa}
{Simmerer} J., {Sneden} C., {Cowan} J.~J., {Collier} J., {Woolf} V.~M.,
  {Lawler} J.~E., 2004, \apj, 617, 1091

\bibitem[{{Sneden}(1973)}]{Sneden:1973aa}
{Sneden} C., 1973, \apj, 184, 839

\bibitem[{{Sneden}, {Cowan} \& {Gallino}(2008){Sneden}, {Cowan}, \&
  {Gallino}}]{Sneden:2008aa}
{Sneden} C., {Cowan} J.~J., {Gallino} R., 2008, \araa, 46, 241

\bibitem[{{Stephens} \& {Boesgaard}(2002)}]{Stephens:2002aa}
{Stephens} A., {Boesgaard} A.~M., 2002, \aj, 123, 1647

\bibitem[{{Tinsley}(1979)}]{Tinsley:1979aa}
{Tinsley} B.~M., 1979, \apj, 229, 1046

\bibitem[{{Tolstoy}, {Hill} \& {Tosi}(2009){Tolstoy}, {Hill}, \&
  {Tosi}}]{Tolstoy:2009aa}
{Tolstoy} E., {Hill} V., {Tosi} M., 2009, \araa, 47, 371

\bibitem[{{Tolstoy} {et~al}\mbox{.}(2003){Tolstoy}, {Venn}, {Shetrone},
  {Primas}, {Hill}, {Kaufer}, \& {Szeifert}}]{Tolstoy:2003aa}
{Tolstoy} E., {Venn} K.~A., {Shetrone} M., {Primas} F., {Hill} V., {Kaufer} A.,
  {Szeifert} T., 2003, \aj, 125, 707

\bibitem[{{Travaglio} {et~al}\mbox{.}(1999){Travaglio}, {Galli}, {Gallino},
  {Busso}, {Ferrini}, \& {Straniero}}]{Travaglio:1999aa}
{Travaglio} C., {Galli} D., {Gallino} R., {Busso} M., {Ferrini} F., {Straniero}
  O., 1999, \apj, 521, 691

\bibitem[{{Trippella} {et~al}\mbox{.}(2016){Trippella}, {Busso}, {Palmerini},
  {Maiorca}, \& {Nucci}}]{Trippella:2016aa}
{Trippella} O., {Busso} M., {Palmerini} S., {Maiorca} E., {Nucci} M.~C., 2016,
  \apj, 818, 125

\bibitem[{{Venn} {et~al}\mbox{.}(2004){Venn}, {Irwin}, {Shetrone}, {Tout},
  {Hill}, \& {Tolstoy}}]{Venn:2004aa}
{Venn} K.~A., {Irwin} M., {Shetrone} M.~D., {Tout} C.~A., {Hill} V., {Tolstoy}
  E., 2004, \aj, 128, 1177

\bibitem[{{Venn} {et~al}\mbox{.}(2012){Venn}, {Shetrone}, {Irwin}, {Hill},
  {Jablonka}, {Tolstoy}, {Lemasle}, {Divell}, {Starkenburg}, {Letarte},
  {Baldner}, {Battaglia}, {Helmi}, {Kaufer}, \& {Primas}}]{Venn:2012aa}
{Venn} K.~A. {et~al.}, 2012, \apj, 751, 102

\bibitem[{{Wallerstein} {et~al}\mbox{.}(1997){Wallerstein}, {Iben}, {Parker},
  {Boesgaard}, {Hale}, {Champagne}, {Barnes}, {K{\"a}ppeler}, {Smith},
  {Hoffman}, {Timmes}, {Sneden}, {Boyd}, {Meyer}, \&
  {Lambert}}]{Wallerstein:1997aa}
{Wallerstein} G. {et~al.}, 1997, Reviews of Modern Physics, 69, 995

\bibitem[{{Wheeler}, {Sneden} \& {Truran}(1989){Wheeler}, {Sneden}, \&
  {Truran}}]{Wheeler:1989aa}
{Wheeler} J.~C., {Sneden} C., {Truran}, Jr. J.~W., 1989, \araa, 27, 279

\bibitem[{{Winckler} {et~al}\mbox{.}(2006){Winckler}, {Dababneh}, {Heil},
  {K{\"a}ppeler}, {Gallino}, \& {Pignatari}}]{Winckler:2006aa}
{Winckler} N., {Dababneh} S., {Heil} M., {K{\"a}ppeler} F., {Gallino} R.,
  {Pignatari} M., 2006, \apj, 647, 685

\bibitem[{{Woosley}, {Heger} \& {Weaver}(2002){Woosley}, {Heger}, \&
  {Weaver}}]{Woosley:2002aa}
{Woosley} S.~E., {Heger} A., {Weaver} T.~A., 2002, Reviews of Modern Physics,
  74, 1015

\bibitem[{{Woosley} \& {Weaver}(1995)}]{Woosley:1995aa}
{Woosley} S.~E., {Weaver} T.~A., 1995, \apjs, 101, 181

\bibitem[{{Yong} {et~al}\mbox{.}(2005){Yong}, {Grundahl}, {Nissen}, {Jensen},
  \& {Lambert}}]{Yong:2005aa}
{Yong} D., {Grundahl} F., {Nissen} P.~E., {Jensen} H.~R., {Lambert} D.~L.,
  2005, \aap, 438, 875

\end{thebibliography}
\end{document}